\documentclass[aps,prstab,superscriptaddress,longbibliography,twocolumn,amsmath,amssymb,floatfix]{revtex4-2}
\usepackage{graphicx}
\usepackage{dcolumn}
\usepackage{hyperref}

\hypersetup{
    colorlinks=true,
    linkcolor=blue,
    citecolor=blue,
    filecolor=magenta,
    urlcolor=blue
}
\newcommand{\Figref}[1]{Fig.\ref{#1}}
\newcommand{\Tabref}[1]{Table~\ref{#1}}

\begin{document}

\title{START-TO-END SIMULATIONS OF NON-RELATIVISTIC ELECTRON ACCELERATION ALONG REFLECTIVE GRATING STRUCTURE BY FRONT-TILTED LASER PULSE}

\author{I.V. Beznosenko}
\altaffiliation[Corresponding author ]{\\Email address: beznosenko1989@gmail.com (I.V. Beznosenko)}

\author{A.V.~Vasyliev}
\author{S.M.~Kovalov}
\author{R.R.~Kniaziev}
\author{G.V.~Sotnikov}

\address{National Science Center Kharkiv Institute of Physics and Technology \\
1, Akademichna St., Kharkiv, 61108, Ukraine}

\date{\today}

\begin{abstract}
A comprehensive (start-to-end) simulation was performed regarding the formation of a front-tilted laser pulse and the subsequent acceleration of non-relativistic electrons moving along a reflective grating structure. The laser pulse tilt was formed using another diffraction grating -- either reflective or transmissive. The pulse had a spectral range (FWHM) of 803--863 nm and an initial duration of approximately 17 fs. Acceleration under normal laser pulse incidence was also calculated for comparison.

Research results indicate that the use of a femtosecond laser pulse with a front, tilted by means of a reflective diffraction grating, significantly enhances the efficiency of laser acceleration of non-relativistic electrons compared to the case of a pulse at normal incidence. Furthermore, it was found that using a transmissive diffraction grating to shape the tilted pulse front also increases acceleration efficiency relative to normal incidence on the accelerating grating, although the acceleration rate is lower than when a reflective diffraction grating is used.

\par PACS: 41.75.Jv, 41.75.Ht, 42.25.Bs
\end{abstract}

\maketitle

\section{INTRODUCTION}
Dielectric laser accelerators (DLAs) are capable of accelerating electrons to energies in the TeV range \cite{England2022}. The theoretical study of laser-driven electron acceleration in a vacuum using femtosecond pulses with various parameters is a topic of current interest in the field of novel particle acceleration methods \cite{Ghota2026}. Employing flat transparent structures to modify the electric field distribution of the laser radiation along the electron path enables more efficient acceleration than using grating-based dielectric structures \cite{Sotnikov2025}. However, the method of electron acceleration using flat transparent structures is applicable only to electrons with energies exceeding a certain threshold. This limit is determined by the refractive index of the material from which the flat structure is fabricated. For instance, if the structure is made of fused silica, the minimum relative electron velocity is 0.688 \cite{Bolshov2021}, corresponding to a minimum energy of 193 keV.

\begin{figure*}[p]
  \centering
  \includegraphics[width=\textwidth]{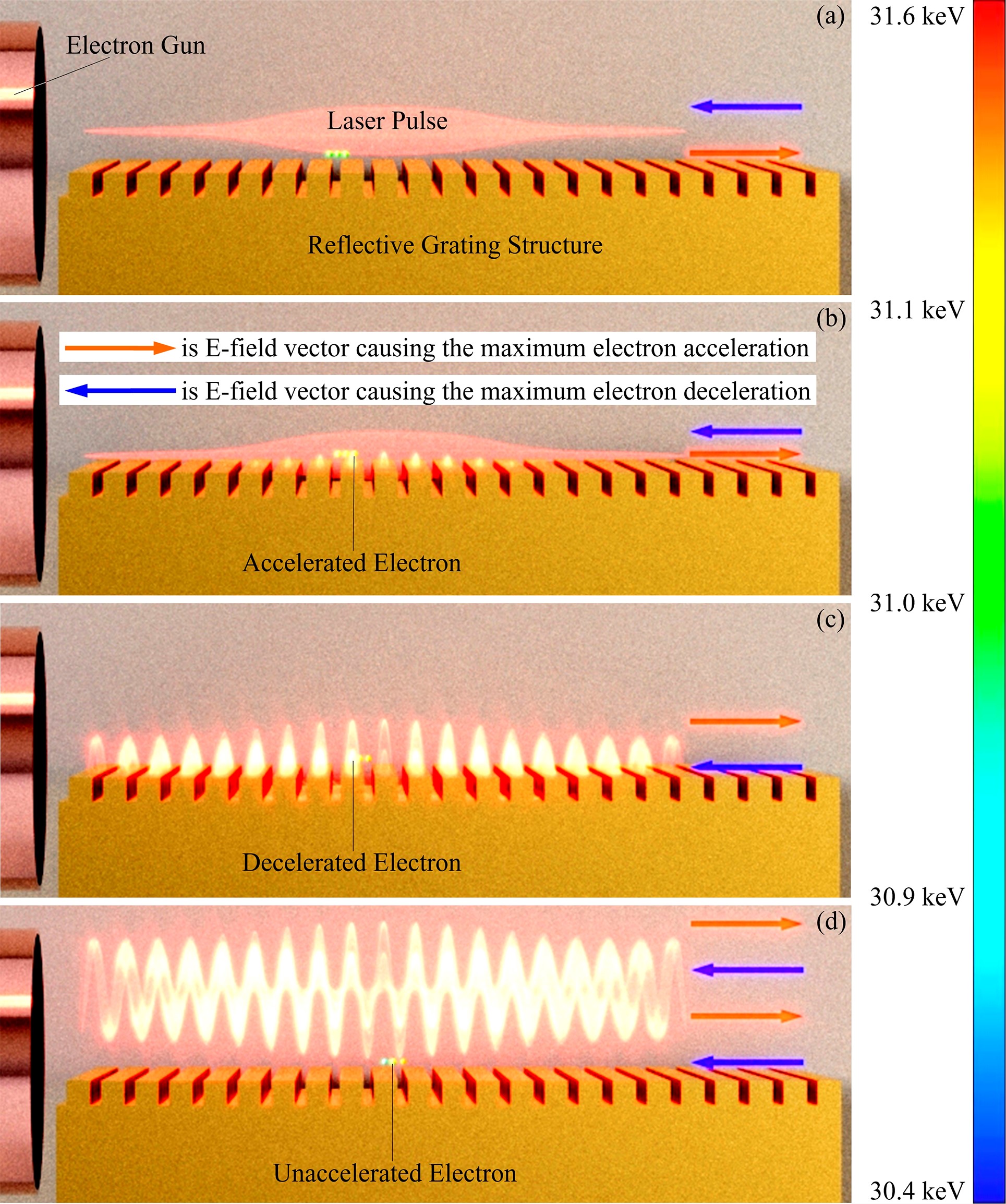}
  \caption{Schematic representation of the use of a laser pulse without front tilting in a laser accelerator based on a reflective grating structure: the moment electrons enter the field of the femtosecond laser pulse (a); the moments when electrons are close to the center of the incident (b) and reflected (c) laser pulse; the moment electrons exit the laser pulse field (d).}
  \label{Fig:01}
\end{figure*}

To accelerate electrons to energies exceeding the limits of conventional electron guns -- reaching the range where flat transparent structures can be employed -- it is advantageous to use Dielectric Laser Accelerators (DLAs) based on various grating-like dielectric structures \cite{Breuer2013,Yousefi2019,Beznosenko2024}, avoiding the need for bulky, conventional radio-frequency accelerators or complex plasma wakefield accelerators. Although electrons interacting with laser radiation along grating structures undergo acceleration with a high peak rate (exceeding 150 MeV/m in optimal DLA configurations), direct acceleration using grating structures suffers from a significant drawback when the laser pulse width greatly exceeds its length. The duration -- and consequently the length -- of the laser pulses used in the femtosecond systems typically employed for DLAs is very small compared to the pulse width along the axis of electron motion. As a result, when such a pulse is incident normally on the accelerating structure, its energy is utilized inefficiently, even when reflective grating structure is used (\Figref{Fig:01}).

\onecolumngrid
\clearpage
\twocolumngrid

In laser accelerators based on grating structures, electron acceleration is driven by the electric component of the laser field, which is aligned with the direction of electron motion. \Figref{Fig:01} illustrates electron dynamics within the field of a laser pulse without front-tilting. The laser pulse is depicted schematically, showing only the high-intensity region; it propagates from top to bottom and, following reflection from the grating structure, propagates from bottom to top in a distorted form. Electrons generated by an electron gun move from left to right. An electron undergoes acceleration if it enters the accelerating phase of the laser field -- meaning the electric field component aligns with the electron's direction of motion -- while positioned above the grating pillars (\Figref{Fig:01},b). As such an accelerated electron passes over the grating grooves, it enters the field's decelerating phase. However, in this region, the electric field component -- which opposes the electron's motion -- is less intense due to the superposition of the field reflected from the groove bottoms (which is in the opposite phase) and diffraction effects. A decelerating electron enters the decelerating phase while above the pillars (\Figref{Fig:01},c). An electron that encounters the neutral phase above both pillars and grooves, and enters the accelerating and decelerating phases only when positioned above the pillar edges (\Figref{Fig:01},d) \cite{Beznosenko2024}, undergoes no significant acceleration or deceleration. Only three electrons are shown in the bunch: the rightmost experiences the greatest acceleration, the leftmost experiences the greatest deceleration, while the kinetic energy of the central electron remains virtually unchanged. The scales and proportions of the elements in the figure have been altered for clarity.

Reference \cite{Cesar2018} proposes an optical scheme for tilting the laser pulse front relative to the electron propagation trajectory using a reflective diffraction grating. In this configuration, the phase front of the laser radiation remains within the plane containing the propagation axis of the accelerated electrons. References \cite{Wei2017,Cesar2018b,Crisp2024,Kozak2018} experimentally confirm the superior efficiency of dielectric accelerators driven by a front-tilted laser pulse compared to a normally incident one.

However, studies \cite{Wei2017,Cesar2018b,Crisp2024} focused solely on the acceleration of relativistic electrons; the electric field of the laser radiation -- both before and after interaction with the diffraction grating -- was not investigated (including in \cite{Kozak2018}). In the simulations, the formation of the laser pulse front tilt and the subsequent electron acceleration were treated as separate processes. Furthermore, the potential for enhancing the efficiency of laser acceleration by using a transmissive grating to tilt the laser pulse front was not considered. Research results \cite{Svystunov2025} indicate that while transmissive dielectric grating structures are advantageous for achieving higher acceleration rates for relativistic electrons, reflective structures \cite{Beznosenko2024} provide more efficient acceleration for non-relativistic electrons.

Thus, comprehensive theoretical studies of the acceleration of non-relativistic electrons along a reflective grating structure using a laser pulse -- whose front is tilted via transmissive and reflective diffraction gratings -- are crucial for selecting an optimal scheme that ensures efficient acceleration during subsequent physical implementation.

\section{PROBLEM FORMULATION}
In the process of calculating a laser-driven accelerator based on front-tilted pulses, the input data consisted of laser radiation and electron beam parameters obtained at the National Science Center ``Kharkiv Institute of Physics and Technology'' (NSC KIPT) -- specifically, at the Institute of Plasma Electronics and New Methods of Acceleration (Ukraine).

A laser pulse was generated using a femtosecond laser system; it has a central wavelength of $\lambda_0 = 833$ nm, a minimum wavelength at half-maximum of 803 nm, and a maximum of 863 nm. Its pulse duration at half-maximum is $t_{\mathrm{FWHM}} \approx 17$ fs \cite{Beznosenko2025}, which corresponds to an FWHM $\approx 5$ $\mu$m \cite{Ahmanov1988}:

\begin{equation}\label{eq:01}
\tau = \frac{4\cdot\ln(a)}{2\cdot\pi\cdot c\cdot(1 / \lambda_{\mathrm{min}} - 1 / \lambda_{\mathrm{max}})} = \frac{l}{c},
\end{equation}

where $\tau$ is laser pulse duration at the $1/a$ level, $c$ is speed of light in air, $\lambda_{\mathrm{min}}$ and $\lambda_{\mathrm{max}}$ are boundaries of the laser pulse emission spectrum at the $1/a$ level, $l$ is pulse duration at the same level ($l = \mathrm{FWHM}$ at $a = 2$), $a = 2$ for the half-maximum.

Using an electron gun and its control unit, stable electron beams with an energy of $E_{k} = 31$ keV were obtained. For the simplest and most efficient laser acceleration of non-relativistic electrons under single-sided irradiation, it is advantageous to use a reflective grating structure (\Figref{Fig:02},a) with a rectangular profile. The height of the structure's pillars is $h = \lambda_{0} / 4 = 208.25$ nm; the initial period of the structure is $\lambda_{\mathrm{p}} = N \cdot \beta \cdot \lambda_{0} = 277.678$ nm, where the relative electron velocity is $\beta = 0.33335$, and the pillars occupy 70\% of the structure's period \cite{Beznosenko2024}. The value $N = 1$ -- representing the number of laser radiation wavelengths traversed by the electron as it passes over a single period of the grating structure -- yields the highest acceleration rate.

A model of this grating structure was used in computer simulations of the laser acceleration of non-relativistic electrons. The electric field vector of the linearly polarized laser pulse accelerating the electrons was oriented parallel to the structure's surface. The distance from the electron beam trajectories to the surface of the grating structure's pillars was set at 150 nm. Electrons were injected throughout the entire duration of the laser radiation's passage across their trajectories. The electron beam diameter was set at 10 nm, and the length of the grating structure was 100 $\mu$m. The laser radiation was directed at the center of the structure; its beam waist radius was set at 50 $\mu$m, while the distance to the grating structure and the focal length were both 300 $\mu$m.

The calculation employed the particle-in-cell (PIC) solver from the Particle Dynamics Simulation package of CST Studio Suite \cite{CST2026}. The formation of the laser pulse front tilt using diffraction gratings and the subsequent acceleration of electrons along the grating structure were performed within a single simulation project.

\section{THEORY OF ELECTRON ACCELERATION USING A FRONT-TILTED LASER PULSE AND A GRATING STRUCTURE}
\begin{figure}[htbp]
  \centering
  \includegraphics[width=0.48\textwidth]{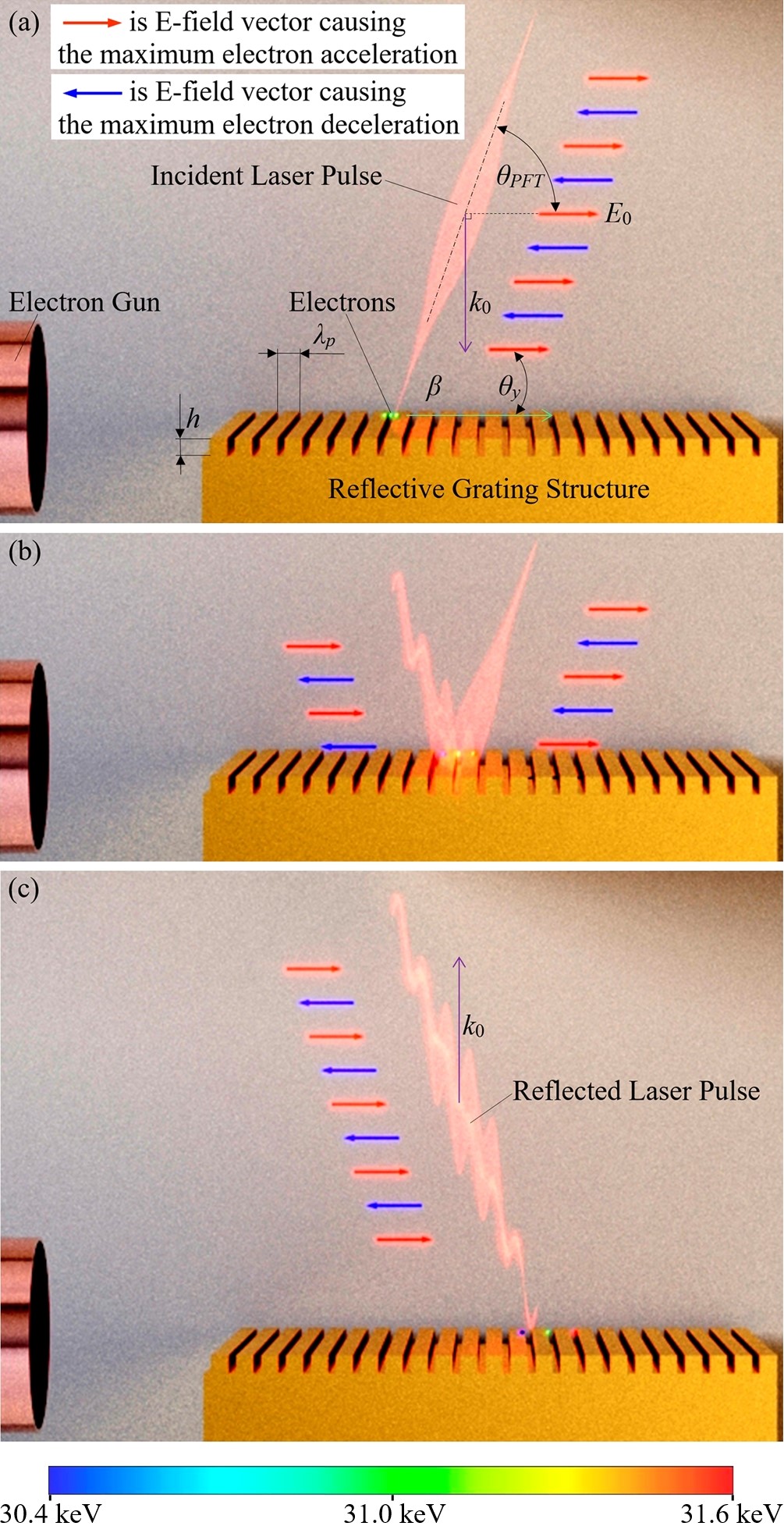}
  \caption{Schematic representation of the use of a front-tilted laser pulse in an accelerator based on a reflective grating structure: the moment electrons enter the field of the femtosecond laser pulse (a); the moment when electrons are close to the center of the incident and reflected laser pulse (b); the moment electrons exit the laser pulse field (c).}
  \label{Fig:02}
\end{figure}

To utilize the laser radiation energy most effectively, two conditions must be met. Firstly, the angle between the E-field vector of the laser radiation and the electron propagation trajectory must be $\theta_y = 0$ (\Figref{Fig:02},a). Secondly, the tilt angle of the laser pulse front relative to its direction of propagation must satisfy condition \cite{Cesar2018b}:

\begin{equation}\label{eq:02}
\theta_{PFT}\approx \arctan (1 / \beta).
\end{equation}

For electrons with an initial energy of $E_{k} \approx 31$ keV, a laser pulse front tilt angle of $\theta_{PFT} = 71^{\circ} 34^{\prime}$ ensures the most efficient acceleration.

\Figref{Fig:02} illustrates electron dynamics during motion within the field of a front-tilted laser pulse. The laser pulse is depicted schematically, showing only the high-intensity region; it propagates from top to bottom and, following reflection from a reflective grating structure, propagates from bottom to top in a distorted form. Electrons generated by an electron gun move from left to right. Only three electrons within the bunch are shown: the rightmost one undergoes the greatest acceleration and the leftmost one undergoes the greatest deceleration, while the kinetic energy of the central electron remains virtually unchanged. The scales and proportions of the elements in the figure have been altered, and the magnitudes of electron acceleration and deceleration have been visually exaggerated for clarity.

It can be seen that the electrons remain in the electric field of the laser radiation for a much longer time when the pulse front is tilted than when there is no tilt (\Figref{Fig:01}).

Such a front tilt can be achieved using diffraction gratings.

\section{ELECTRON ACCELERATION USING A FRONT-TILTED LASER PULSE FORMED BY A TRANSMISSION DIFFRACTION GRATING}
\begin{figure}[htbp]
  \centering
  \includegraphics[width=0.48\textwidth]{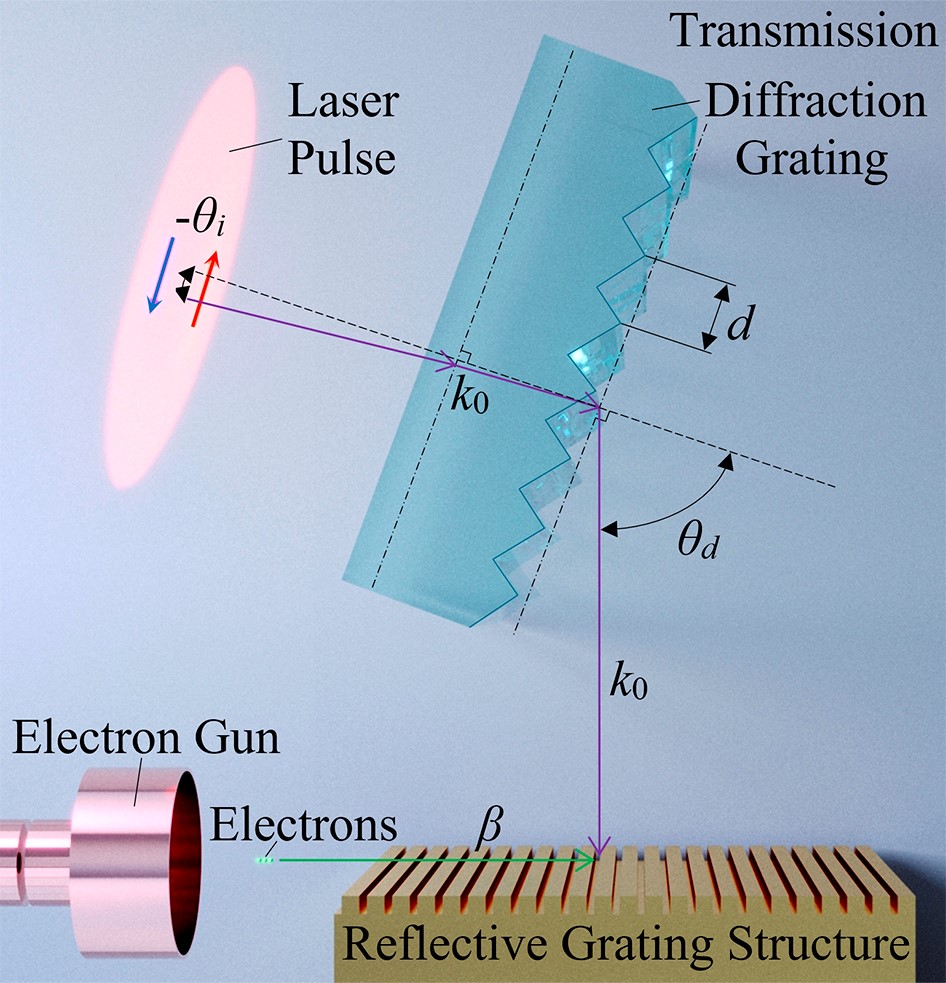}
  \caption{Schematic representation of laser pulse front tilting using a transmissive diffraction grating in an accelerator.}
  \label{Fig:03}
\end{figure}

To effectively accelerate electrons using a front-tilted laser pulse formed by a transmission diffraction grating, it is necessary to ensure precise alignment of this grating relative to the accelerating grating structure, as well as precise alignment of the laser beam relative to the grating. The pulse-front tilt angle $\theta_{PFT}$ (\Figref{Fig:02}) is related to the exit angle $\theta_{d}$ of the first diffraction order of the laser radiation from the transmission diffraction grating (\Figref{Fig:03}) by the following expression \cite{Cesar2018}:

\begin{equation}\label{eq:03}
\theta_{PFT} = \arctan \left(\frac{\lambda_0}{d\cdot\cos(\theta_d)}\right).
\end{equation}

To achieve maximum acceleration of the electron bunch, the E-field vector of the laser radiation at the moment of interaction with the electrons must be directed predominantly along the electron bunch's trajectory. In other words, the laser radiation, having passed through the transmissive diffraction grating, must propagate perpendicularly to both the electron trajectory and the surface of the grating structure -- that is, $\theta_{y} = 0$ (\Figref{Fig:02},a). By combining expressions (\ref{eq:02}) and (\ref{eq:03}), one can express $\theta_{d}$ in terms of the relative electron velocity $\beta$:

\begin{equation}\label{eq:04}
\theta_{d} = \arccos \left(\frac{\beta\cdot\lambda_0}{d}\right).
\end{equation}

The exit angle $\theta_{d}$ of the first-order diffracted radiation from a transmission diffraction grating with period $d$ and the angle of incidence $\theta_{i}$ of the radiation upon it (\Figref{Fig:03}) are related by the following expression \cite{Schiller1980}:

\begin{equation}\label{eq:05}
d(\sin (\theta_{d}) - \sin (\theta_{i})) = \lambda_{0},
\end{equation}

then:

\begin{equation}\label{eq:06}
\theta_{i} = \arcsin (\sin (\theta_{d}) - \lambda_{0} / d).
\end{equation}

Thus, for the efficient acceleration of electrons using short laser pulses, it is necessary to select a transmission diffraction grating such that the angles $\theta_{d}$ (\ref{eq:04}) and $\theta_{i}$ (\ref{eq:06}) are not imaginary. The Thorlabs GT13-12 stock diffraction grating, with a period of $d = 833.3$ nm \cite{Thorlabs2026}, satisfies this requirement; for this grating, $\theta_{d} = 70^{\circ}32^{\prime}$ and $\theta_{i} = -3^{\circ}15^{\prime}$.

A computer simulation of such acceleration was performed. An electron energy gain of 300 eV was obtained with a specified laser E-field amplitude of 6 GV/m -- a value that does not exceed the damage threshold of a fused-silica transmissive diffraction grating \cite{Breuer2013}. Near the accelerating reflective grating structure, the peak E-field was 0.1 GV/m, which does not exceed the damage threshold for such structures under femtosecond pulse irradiation \cite{Breuer2013}.

A computer simulation of the acceleration process was also performed under similar conditions, but for the case of a laser pulse incident without front tilting (\Figref{Fig:01}). An electron energy gain of 705 eV was obtained. The peak E-field near the accelerating grating structure was 1.1 GV/m. Thus, at laser E-field that are close to -- but do not exceed -- the damage thresholds of the target structures, the efficiency of an accelerator utilizing a front-tilted pulse (formed by a transmission diffraction grating) was found to be 2.4 times lower than that of an accelerator using a pulse without front tilt. This is attributed to the significantly larger fraction of zero-order diffracted radiation compared to the first order upon passage through the transmission diffraction grating, as well as to a substantial amount of reflected radiation (\Figref{Fig:05},a).

A computer simulation of similar acceleration was also performed for the case of a laser pulse incident without front tilt, where the peak E-field along the electron bunch axis near the center of the accelerating grating structure was 0.1 GV/m -- the same value used in the simulation of electron acceleration using a front-tilted laser pulse formed by a transmissive diffraction grating. To achieve this, the amplitude of the laser E-field was set to 0.55 GV/m. The resulting electron energy gain was 63 eV. Consequently, for identical peak E-field along the electron path, the efficiency of an accelerator utilizing a front-tilted pulse is 4.8 times higher than that of one using a pulse without front tilt. This ratio increases with the length of the accelerating grating structure and the radius of the laser beam waist. These findings justify the use of transmissive diffraction gratings to tilt the front of short laser pulses during electron acceleration.

\section{ELECTRON ACCELERATION USING A FRONT-TILTED LASER PULSE FORMED BY A REFLECTIVE DIFFRACTION GRATING}
\begin{figure}[htbp]
  \centering
  \includegraphics[width=0.48\textwidth]{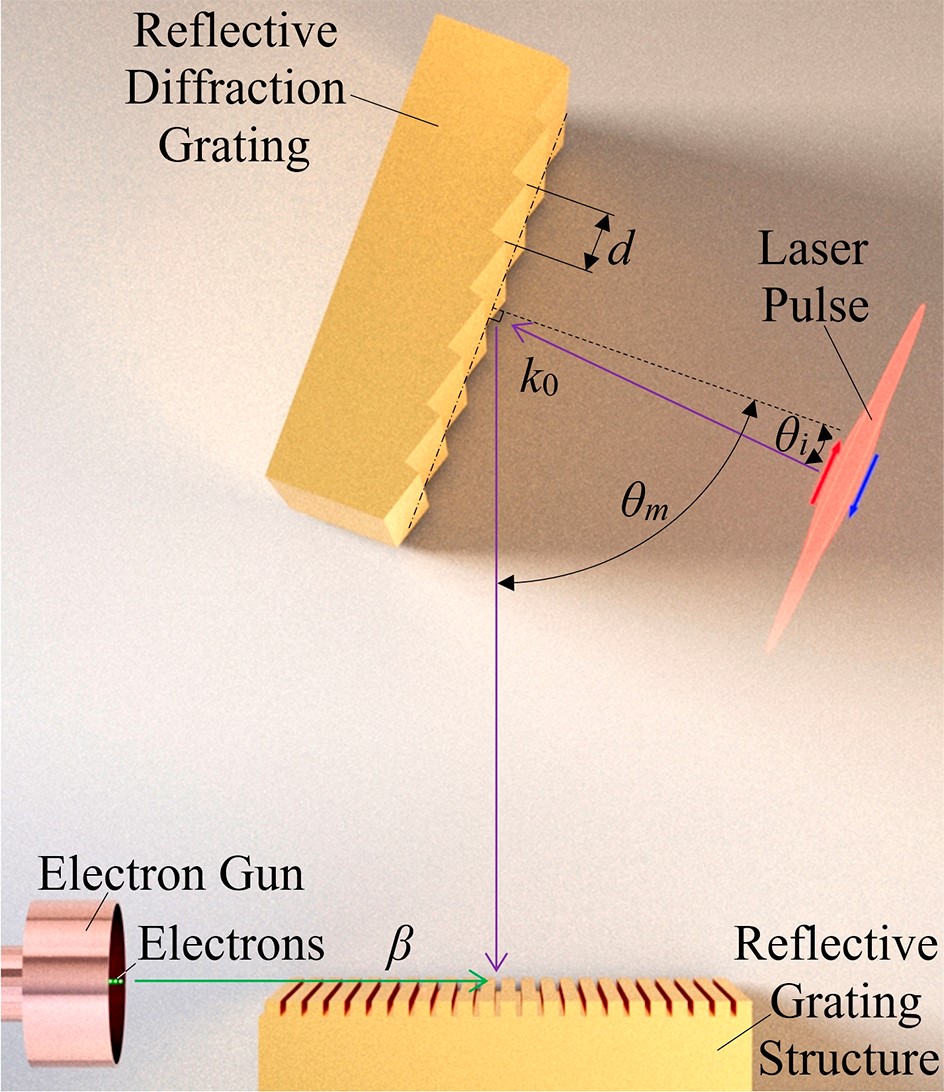}
  \caption{Schematic representation of laser pulse front tilting using a reflective diffraction grating in an accelerator.}
  \label{Fig:04}
\end{figure}

To effectively accelerate electrons using a front-tilted laser pulse formed by a reflective diffraction grating, it is necessary to ensure precise alignment of this grating relative to the accelerating grating structure, as well as precise alignment of the laser beam relative to the grating. The pulse front tilt angle $\theta_{PFT}$ (\Figref{Fig:02}) is related to the reflection angle $\theta_{m}$ of the first diffraction order of the laser radiation from the reflective diffraction grating (\Figref{Fig:04}) by the following expression \cite{Cesar2018}:

\begin{equation}\label{eq:07}
\theta_{PFT} = \arctan \left(\frac{\lambda_0}{d\cdot\cos(\theta_m)}\right).
\end{equation}

To achieve maximum acceleration of the electron bunch, the E-field vector of the laser radiation at the moment of interaction with the electrons must be directed predominantly along the electron bunch's trajectory. In other words, the laser radiation reflected from the reflective diffraction grating must propagate perpendicularly to both the electron trajectory and the surface of the grating structure -- that is, $\theta_{y} = 0$ (\Figref{Fig:02},a). By combining expressions (\ref{eq:02}) and (\ref{eq:07}), one can express $\theta_{m}$ in terms of the relative electron velocity $\beta$:

\begin{equation}\label{eq:08}
\theta_{m} = \arccos \left(\frac{\beta\cdot\lambda_{0}}{d}\right).
\end{equation}

The angle of reflection $\theta_{m}$ of the first-order diffracted radiation from a reflective diffraction grating with period $d$ and the angle of incidence $\theta_{i}$ of the radiation upon it (\Figref{Fig:04}) are related by the following expression \cite{Schiller1980}:

\begin{equation}\label{eq:09}
d\cdot (\sin (\theta_{m}) + \sin (\theta_{i})) = \lambda_{0},
\end{equation}

then:

\begin{equation}\label{eq:10}
\theta_{i} = \arcsin (\lambda_{0} / d - \sin (\theta_{m})).
\end{equation}

Thus, for the efficient acceleration of electrons using short laser pulses, it is necessary to select a reflective diffraction grating such that the angles $\theta_{m}$ (\ref{eq:08}) and $\theta_{i}$ (\ref{eq:10}) are not imaginary. The ThorLabs GR13-1205 stock diffraction grating, with a period of $d = 833.3$ nm \cite{Thorlabs2026}, satisfies this requirement; for this grating, $\theta_{m} = 70^{\circ}32^{\prime}$ and $\theta_{i} = 3^{\circ}15^{\prime}$.

A computer simulation of such acceleration was performed. An electron energy gain of 600 eV was obtained at a specified laser E-field amplitude of 1.8 GV/m -- this value does not exceed the damage threshold of a gold-coated reflective diffraction grating -- a threshold three times lower than that of an uncoated fused-silica grating \cite{Breuer2013}. Near the accelerating reflective grating structure, the peak E-field was 0.13 GV/m, which does not exceed the damage threshold for such structures under femtosecond pulse irradiation.

A computer simulation of the acceleration process was also performed under similar conditions but for the case of a laser pulse incident without front tilt (\Figref{Fig:01}). An electron energy gain of 205 eV was obtained. The peak E-field near the accelerating grating structure was 0.33 GV/m. Thus, for identical laser E-field amplitudes chosen to be close to, yet below, the damage thresholds of the target structures -- the efficiency of an accelerator utilizing a front-tilted pulse (formed by a reflective diffraction grating) was found to be 2.9 times higher than that of an accelerator using a pulse without front tilt.

A computer simulation of similar acceleration was also performed for the case of a laser pulse incident without a front tilt, where the peak E-field along the electron bunch axis near the center of the accelerating grating structure was 0.13 GV/m -- the same value used in the simulation of electron acceleration using a front-tilted laser pulse formed by a reflective diffraction grating. To achieve this, the amplitude of the laser E-field was set to 0.709 GV/m. The resulting electron energy gain was 80 eV. Consequently, for identical peak E-field along the electron trajectory, the efficiency of an accelerator utilizing a front-tilted pulse is 7.5 times higher than that of one using a pulse without a front tilt. This ratio increases with the length of the accelerating grating structure and the radius of the laser beam waist.

This justifies the advisability of using reflective diffraction gratings to tilt the front of short laser pulses during electron acceleration.

\section{COMPARISON OF ELECTRON ACCELERATION USING A FRONT-TILTED LASER PULSE FORMED BY TRANSMISSIVE AND REFLECTIVE DIFFRACTION GRATINGS}
To compare the efficiency of electron acceleration using a front-tilted laser pulse -- formed by either a transmissive diffraction grating or a reflective one -- additional calculations were performed for the first case at specified laser E-field amplitudes of 1.8 GV/m and 7.8 GV/m; these yielded electron energy gains of 97 eV and 373 eV, respectively.

\begin{figure}[htbp]
  \centering
  \includegraphics[width=0.48\textwidth]{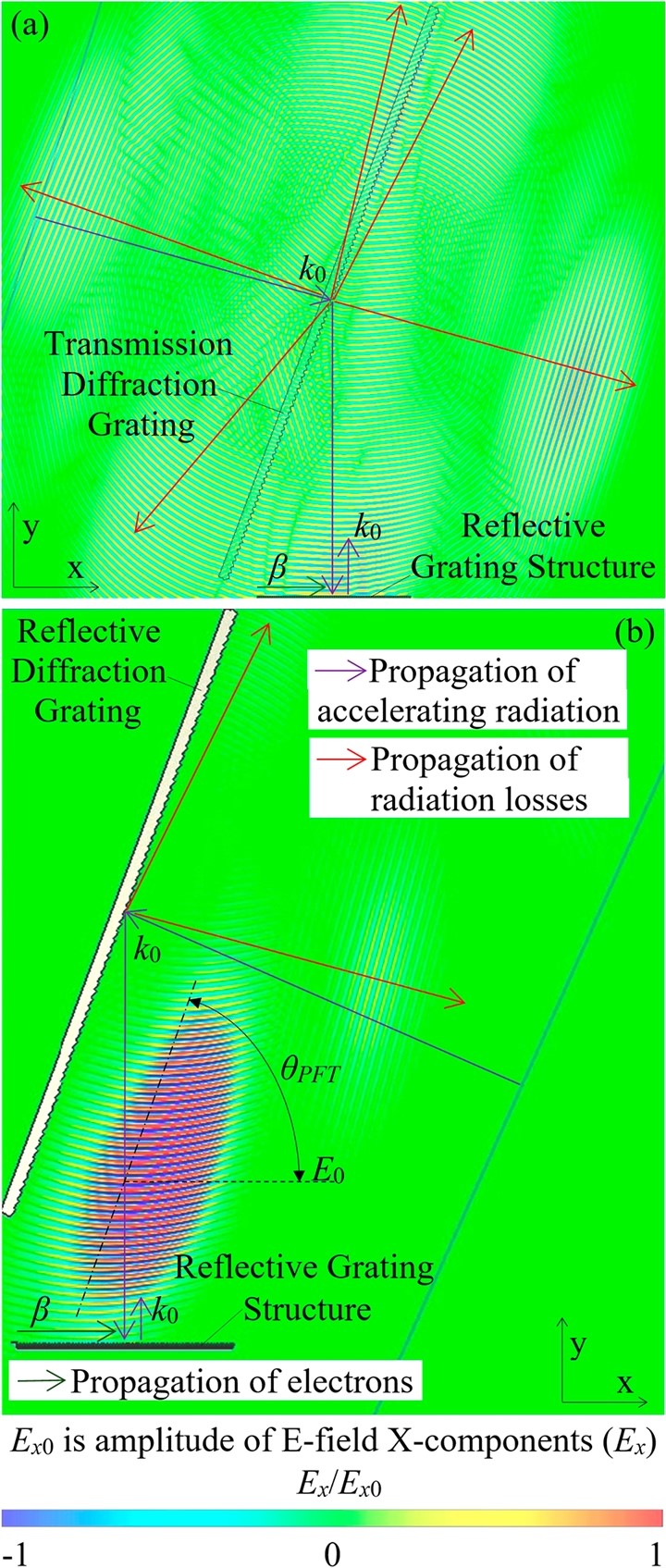}
  \caption{Distribution of the X-component of the E-field for cases where the laser pulse front is tilted using a transmissive (a) and a reflective (b) diffraction grating.}
  \label{Fig:05}
\end{figure}

For identical specified amplitudes of the laser E-field (1.8 GV/m), the use of a reflective diffraction grating yields an energy gain 6.2 times greater than that of a transmissive grating (600 eV versus 97 eV). This is attributed to the significantly larger fraction of zero-order diffracted radiation compared to the first order when passing through a transmissive grating, as well as to the substantial proportion of reflected radiation (\Figref{Fig:05},a).

For the same peak E-field on the electron bunch axis (0.13 GV/m), the use of a reflective diffraction grating yields an energy gain 1.6 times greater than that of a transmissive grating (600 eV versus 373 eV). This is attributed to the distortion of the laser pulse's phase front as it passes through the material of the transmissive diffraction grating, since the portion of the radiation traversing a shorter path through the grating teeth outruns the portion traversing a longer path.

\Figref{Fig:05} shows the calculated qualitative distribution of the X-component of the E-field influencing electron acceleration for cases where the laser pulse front is tilted using transmissive (\Figref{Fig:05},a) and reflective (\Figref{Fig:05},b) diffraction gratings. The purple arrow indicates the propagation direction of the laser pulse before reaching the grating and the subsequent propagation of the first-order diffracted laser radiation, which accelerates electrons moving along the reflective grating structure. Red arrows indicate the propagation directions of laser radiation components corresponding to other diffraction orders, as well as radiation reflected from the transmissive diffraction grating. The green arrow indicates the initial direction of electron motion at a relative velocity $\beta$. In these calculations, the focal length was set to 80 $\mu$m, the beam waist radius near the grating structure to 10 $\mu$m, and the length of the structure itself to 20 $\mu$m. The lengths of the diffraction gratings were also reduced. All other parameters remained unchanged. It can be seen that when a reflective grating is used (\Figref{Fig:05},b), the fraction of first-order diffracted radiation is significantly larger than the zero-order fraction, in contrast to the case where a transmissive grating is used (\Figref{Fig:05},a).

It can be concluded that the use of reflective diffraction gratings for front tilting of short laser pulses in a laser-driven accelerator will yield the greatest increase in the energy of accelerated electrons at relatively low laser pulse intensities.

\section{CONCLUSIONS}
A comparative analysis of the electron energy gain achieved in the laser accelerator demonstrated that using a reflective diffraction grating to tilt the femtosecond laser pulse front significantly enhances acceleration efficiency compared to the case of normal incidence on the accelerating structure. This is shown in \Tabref{Tabl_1}.

\begin{table}
   \centering
   \caption{Energy gain of electrons with an initial energy of 31 keV following laser acceleration at a specified laser E-field amplitude of 1.8 GV/m}
   \begin{tabular}{lcc}
   \hline
   \hline
       \textbf{Type of laser} & \textbf{Electron} & \textbf{Peak E-field} \\
       \textbf{acceleration} & \textbf{energy gain} & \textbf{near the} \\
       \textbf{of electrons} & \textbf{(eV)} & \textbf{accelerating} \\
       & & \textbf{grating} \\
       & & \textbf{structure} \\
       & & \textbf{(GV/m)} \\
   \hline
   \hline
          Normally-incident & 205 & 0.33 \\
          laser pulse acceleration & & \\
   \hline
          Front-tilted ($\theta_{PFT}$ = 71$^{\circ}$ 34$^{\prime}$) & 600 & 0.13 \\
          laser pulse acceleration & & \\
          using reflective & & \\
          diffraction grating & & \\
   \hline
          Front-tilted ($\theta_{PFT}$ = 71$^{\circ}$ 34$^{\prime}$) & 97 & 0.03 \\
          laser pulse acceleration & & \\
          using transmissive & & \\
          diffraction grating & & \\
   \hline
   \hline
   \end{tabular}
   \label{Tabl_1}
\end{table}

The use of a transmission diffraction grating also significantly increases the efficiency of the laser accelerator, but only when the peak E-field of the laser radiation near the accelerating grating structure are identical -- that is, when the specified amplitude is increased. This is shown in \Tabref{Tabl_2}.

\begin{table}
   \centering
   \caption{Energy gain of electrons with an initial energy of 31 keV following laser acceleration at various specified amplitudes of the laser E-field}
   \begin{tabular}{lccc}
   \hline
   \hline
       \textbf{Type of laser} & \textbf{Electron} & \textbf{Specified} & \textbf{Peak E-field} \\
       \textbf{acceleration} & \textbf{energy gain} & \textbf{E-field} & \textbf{near the} \\
       \textbf{of electrons} & \textbf{(eV)} & \textbf{amplitude} & \textbf{accelerating} \\
       & & \textbf{of laser} & \textbf{grating} \\
       & & \textbf{radiation} & \textbf{structure} \\
       & & \textbf{(GV/m)} & \textbf{(GV/m)} \\
   \hline
   \hline
          Normally-incident & 63 & 0.55 & 0.10 \\
          laser pulse & 705 & 6.00 & 1.10 \\
          acceleration & 80 & 0.71 & 0.13 \\
   \hline
          Front-tilted & 373 & 7.80 & 0.13 \\
          ($\theta_{PFT}$ = 71$^{\circ}$ 34$^{\prime}$) & 300 & 6.00 & 0.10 \\
          laser pulse & & & \\
          acceleration using & & & \\
          transmissive & & & \\
          diffraction grating & & & \\
   \hline
   \hline
   \end{tabular}
   \label{Tabl_2}
\end{table}

The results demonstrate that laser pulse front tilting using diffraction gratings is a promising method for enhancing the efficiency of laser accelerators of non-relativistic electrons moving along a reflective grating structure.

\begin{figure}[htbp]
  \centering
  \includegraphics[width=0.48\textwidth]{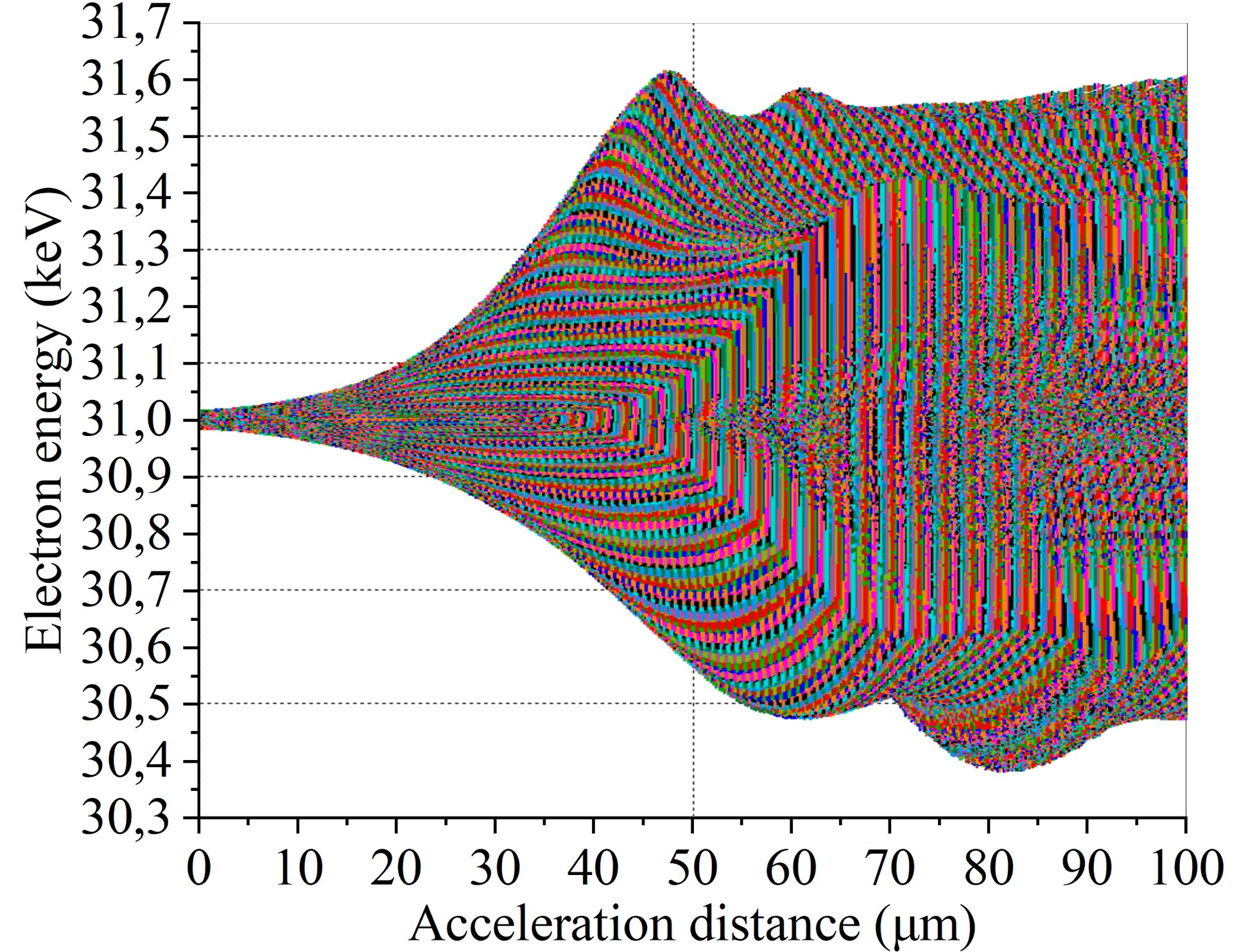}
  \caption{Dependence of electron energy on the path passed over an accelerating grating structure during acceleration by a front-tilted laser pulse formed by a reflective diffraction grating.}
  \label{Fig:06}
\end{figure}

Studies have also shown that, to increase the rate of electron acceleration in such laser accelerators, a number of challenges must be addressed:

\begin{enumerate}
\item optimization of the computational model to prevent a significant drop in the peak E-field of the laser pulse during propagation simulations -- which can be achieved by selecting a pulse with a longer duration and reducing its propagation path;

\item optimization of the period of the accelerating grating structure, since the dependence of electron energy on the path passed over this structure during acceleration is asymmetric, and the energy gain ceases as early as the center of the structure (\Figref{Fig:06});

\item use of bilateral electron irradiation instead of unilateral, and a dual accelerating structure instead of a single one, since the dependence of electron energy on the path passed over the structure during acceleration is discontinuous (\Figref{Fig:06}); this indicates that some electrons strike the accelerating structure due to significant lateral deflection of their motion, caused by the asymmetry of the laser E-field along the electron trajectory.
\end{enumerate}

\begin{acknowledgments}
The study is supported by the National Research Foundation of Ukraine under the program ``Excellent Science in Ukraine'' (project \# 2023.03/0182).
\end{acknowledgments}

\end{document}